\documentclass[twocolumn]{article}

\usepackage{arxiv}

\usepackage{url,hyperref,lineno,microtype}
\usepackage[onehalfspacing]{setspace}
\usepackage[english]{babel}

\usepackage{amsmath,amssymb,amsfonts}

\usepackage{tabularx}
\usepackage{booktabs}
\usepackage[usenames,dvipsnames]{xcolor}
\usepackage{colortbl}
\definecolor{Gray}{gray}{0.925}

\usepackage{qtree}
\usepackage{float}

\usepackage{xspace}
\makeatletter
\DeclareRobustCommand\onedot{\futurelet\@let@token\@onedot}
\def\@onedot{\ifx\@let@token.\else.\null\fi\xspace}
\def\eg{e.g\onedot} 
\def\ie{i.e\onedot} \def\Ie{\emph{I.e}\onedot}
 
\def\etc{etc\onedot}

\makeatother

\usepackage{etoolbox}
\expandafter\patchcmd\csname citeauthor \endcsname
  {\begingroup}{\begingroup\aftergroup\@}{}{}

\usepackage{float}
\usepackage{pdfpages}
\usepackage{tikz}
\usetikzlibrary{automata,positioning}

\usepackage{algorithmic}
\usepackage{graphicx}
\usepackage{textcomp}

\hypersetup{
    colorlinks,
    linkcolor={red!50!black},
    citecolor={blue!50!black},
    urlcolor={blue!80!black}
}

\usepackage{cuted}

\usepackage[capitalize, noabbrev]{cleveref}
\usepackage{natbib}
\def\BibTeX{{\rm B\kern-.05em{\sc i\kern-.025em b}\kern-.08em
    T\kern-.1667em\lower.7ex\hbox{E}\kern-.125emX}}

\begin{document}

\title{A Systematic Review on Model Watermarking for Neural Networks} 
\author{
  Franziska Boenisch\\
  Fraunhofer AISEC and Freie University, Berlin, Germany\\
  \texttt{franziska.boenisch@aisec.fraunhofer.de} \\
}

\maketitle
\begin{strip}
\vspace{16ex}
\end{strip}
\vfill\break

\begin{abstract}
Machine learning (ML) models are applied in an increasing variety of domains. 
The availability of large amounts of data and computational resources encourages the development of ever more complex and valuable models. 
These models are considered intellectual property of the legitimate parties who have trained them, which makes their protection against stealing, illegitimate redistribution, and unauthorized application an urgent need. 
Digital watermarking presents a strong mechanism for marking model ownership and, thereby, offers protection against those threats. 
This work presents a taxonomy identifying and analyzing different classes of watermarking schemes for ML models.
It introduces a unified threat model to allow structured reasoning on and comparison of the effectiveness of watermarking methods in different scenarios.
Furthermore, it systematizes desired security requirements and attacks against ML model watermarking.
Based on that framework, representative literature from the field is surveyed to illustrate the taxonomy.
Finally, shortcomings and general limitations of existing approaches are discussed, and an outlook on future research directions is given.

\end{abstract}

\section{Introduction}
\label{sec:introduction}
In recent years, machine learning (ML) has experienced great advancements. 
Its ability to process ever larger and more complex datasets has led to its application in a versatile and growing number of domains.
The performance of the applied models, thereby, largely depends on the quality and quantity of their training data.
However, the process of training data collection, cleansing, processing, organizing, storing, and in certain cases even manual labeling is time-consuming and expensive.
So is the training process itself, as it may require large computational capacities, \eg in the form of numerous GPUs, and know-how for hyperparameter-tuning.
As a consequence, a trained ML model may be of high value and is to be considered intellectual property of the legitimate owner, \ie the party that created it.

The value incorporated in trained ML models may turn them into lucrative attack targets for malicious attackers who want to steal their functionality \citep{Ateniese.2013Hacking} for redistribution or to offer their own paid services based on them.
Given the broad attack surface of stealing ML models, it might be impossible to entirely prevent theft.
If theft cannot be prevented beforehand, a legitimate model owner might want to react, at least, to the inflicted damage and claim copyright to take further steps.
This, however, requires that the stolen intellectual property can be traced back to its legitimate owner through adequate labeling. 

\begin{figure}[t]
  \centering
  \includegraphics[width=.45\textwidth]{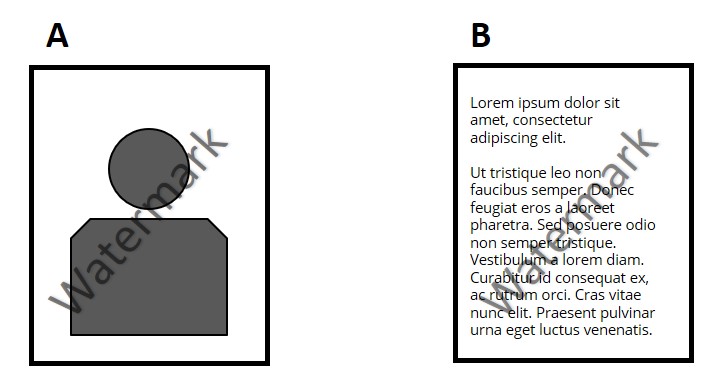}
\caption{Example of a digital watermark embedded in (A) an image and (B) a text file.}
\label{fig:examplewatermark}
\end{figure}

The idea of marking digital property is called \emph{watermarking}.
It refers to the act of embedding identification information into some original data to claim copyright, however, without affecting the data usage.
Watermarking is already broadly applied in digital media, for example in images, where a watermark may consist of a company logo inserted somewhere into the picture, or in texts, where a watermark in form of an identifying text or image might be added to the file.
See Figure~\ref{fig:examplewatermark} for an example of such digital watermarks.
Also, see \cite{Saini.2014Survey} for a survey on watermarking approaches in digital media.

The concept of watermarking can also be adopted for tagging ML models.
So far, several methods to generate such watermarks in ML models have been proposed in research.
Additionally, ways to detect, suppress, remove, or forge existing watermarks have been proposed.
However, so far, the threat space in which the watermarking schemes operate has not been properly characterized. 
The same holds for the goals and guarantees offered by the different watermarking approaches. 
This makes it difficult for model owners to choose an adequate watermarking scheme that fulfills the needs in their scenarios, but also to compare existing approaches with each other. 
Those shortcomings present at the same time a motivation and challenge for the systematic review put forward in this work.
The goal of the article is to introduce a unified threat space for model watermarking, as well as a taxonomy of watermarking methods, and a systematization of their requirements.
It, thereby, does not only propose a common language for evaluating NN watermarking, but goes beyond and enables a structured comparison among existing approaches.
This can serve as a basis to make watermarking methods more usable, comparable, and accessible.

The concrete contributions by this work are as follows:
\begin{itemize}
    \item 
    Taxonomy for watermarking schemes.
    \item Systematization of desirable security properties of ML model watermarks and attacks against them.
    \item 
    Introduction of a unified threat model that enables structured analyses of existing watermarking schemes.
    \item Survey and evaluation of existing watermarking schemes and defenses according to the presented properties.
\end{itemize}
Based on the developed threat model, representative and influential works from the literature were selected.
Although attempts were made to provide a comprehensive and complete overview, it is practically not possible to cite all works in the scope of the given article.
For example, the topic of side-channel attacks that aim at extracting neural networks \citep{Wei.2018know} is not covered.
Also, the watermarking schemes presented in the following are limited to neural networks (NN) for classification.

\section{Basic Concepts and Background}
\label{sec:concepts}

This section provides a brief overview on ML, on model stealing in general and model extraction attacks in particular, and introduces the concept of backdoors for NNs.

\subsection{Machine Learning}
ML consists of two phases, \emph{training} and \emph{inference/testing}.
During training, an ML model $h_\theta$ is given a training dataset $\mathcal{D}_s$ to fit its parameters $\theta$ on.
For classification tasks, the training data has the form $(x,y)$ with $x$ denoting the \emph{feature vector} and $y$ the \emph{target class}.
The model parameters are adjusted through minimizing a loss function that expresses the distance between predictions $h_\theta(x)$ and true targets $y$~\citep{Papernot.2018SoK}. 

At test time, once the model parameters $\theta$ are fit, the function $h_\theta()$ can be applied to new and unseen data $x'$ to produce predictions on them.
A model that performs well solely on the training data is called to \emph{overfit} that data, whereas a model that also performs well on the unseen test data is said to exhibit a good \emph{generalization capacity}.
Performance is usually expressed in terms of \emph{accuracy}, which is the percentage of correct predictions over all predictions.

\subsection{Model Stealing}
\begin{figure*}[t]
  \centering
  \includegraphics[width=.8\textwidth]{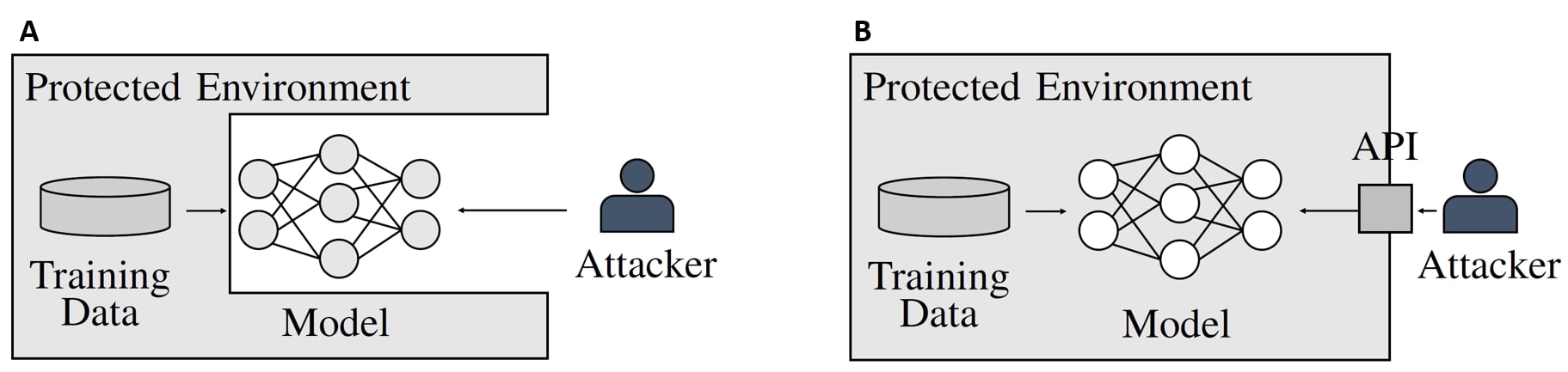}  
\caption{Access scenarios for ML models. (\textbf{A}) A white-box setting allows the attacker full access to the model and all of its parameters, but not (necessarily) to the model's training data. (\textbf{B}) In a black-box scenario, the attacker has no direct access to the model, but instead interacts with it over an application programming interface (API).}
\label{fig:access}
\end{figure*}

Potential attackers may attempt to steal an ML model to have unlimited access to its complex functionality without the high preparation, or continuous per query costs.
Alternatively, they may wish to use the stolen model as a departure point for further attacks that are rendered more efficient through model parameter access, \eg adversarial sample crafting \citep{Tramer.2016Stealing,Carlini.2020Cryptanalytic}.
Protecting ML models against theft is a challenging task, as by definition, the models are supposed to reveal some information to the users.
Hence, in addition to the classical security risks of model theft  \eg malicious insider access, successful attacks on servers hosting the model, or side-channel attacks \citep{Batina.2018CSI}, the information legally revealed by the model can be exploited.
This enables to steal ML models in white-box as well as in black-box settings (see Figure~\ref{fig:access}).

\subsection{Model Extraction}
\label{sub:modelextraction}

\begin{figure}[t]
  \centering
  \includegraphics[width=.45\textwidth]{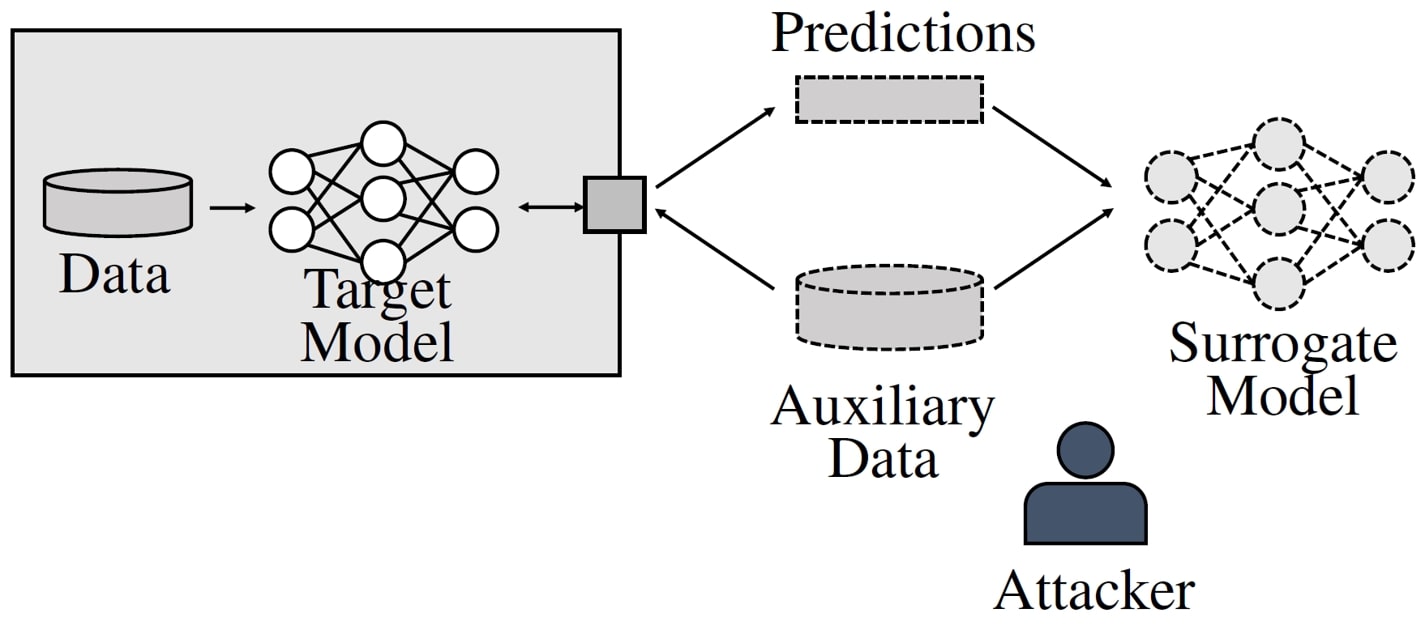}
  \caption{Process of a model extraction attack. The attacker holds auxiliary data from a similar distribution as the target model's training data. Through query access, the attacker obtains corresponding labels for the auxiliary data. Based on that data and the labels, a surrogate model can be trained that exhibits a similar functionality to the original model.}
  \label{fig:stealing}
\end{figure}

A \emph{model extraction} attack refers to stealing a target model $h_\theta$ through black-box access, \ie through posing queries to the model over a predefined interface as depicted in Figure~\ref{fig:access}B.
An attacker might use those queries to $h_\theta$ to obtain labels for unlabeled data $D_{s'}$ from distribution $D$.
Given $D_{s'}$ and the corresponding labels obtained from the original model, the attacker can train a \emph{surrogate model} $h'_\theta$ that incorporates the original model's functionality.
See Figure~\ref{fig:stealing} for an overview on the process. \cite{Jagielski.2019High} distinguish between two types of \emph{model extraction}.
A \emph{fidelity extraction} attack is considered successful, if $h'_\theta$ reproduces $h_\theta$ with small deviation.
Hence, when $h_\theta$ is erroneous w.r.t. the ground truth label of a data point, so is $h'_\theta$.
A \emph{task accuracy extraction} aims at extracting a model that solves approximately the same underlying decision task~\citep{Jagielski.2019High}.

\subsection{Backdoors in NNs}
\label{sub:backdoors}
\cite{Adi.2018Turning} define backdooring in NNs as a technique to intentionally train an ML model to output incorrect predictions (w.r.t. the ground truth) on a given set of the training data points.
As a result, the backdoored NN behaves normally on most data points but differently on the backdoor data \citep{liu2018fine}.
The ability to add backdoors to NNs results from their over-parametrization, \ie the fact that many such models contain more parameters than they need for solving the task that they are supposed to solve. 
In watermarking, a set of backdoor data points can, for example, be used to mark and later recognize a trained NN.
Thereby, the backdoor data can act as a watermark trigger (see Section~\ref{sub:embedding}).

\section{Taxonomy of NN Watermarking}
\label{sec:taxonomy}
Even though inserting a watermark into a model does not prevent theft, it can still enable legitimate owners to identify their stolen model instances.
Therefore, after the model is stolen by an attacker, the legitimate owner might use the watermark to re-identify it and claim copyright. 
Hence, the watermarking methods need to be effective in, and chosen adequately to the given scenario.
For example, using a watermarking scheme that does not offer any binding between the watermark and the identity of the legitimate model owner might still help the owner recognize stolen model instances.
However, it is of little use in a scenario where the owner wants to claim copyright in front of a third party, such as a legal entity.
This section presents a taxonomy to support classifying  watermarking schemes along five different dimensions.
Such a classification can be helpful for identifying and comparing adequate watermarking schemes for a concrete scenario and the corresponding requirements.

\begin{figure*}[ht]
  \centering
  \includegraphics[width=.8\textwidth]{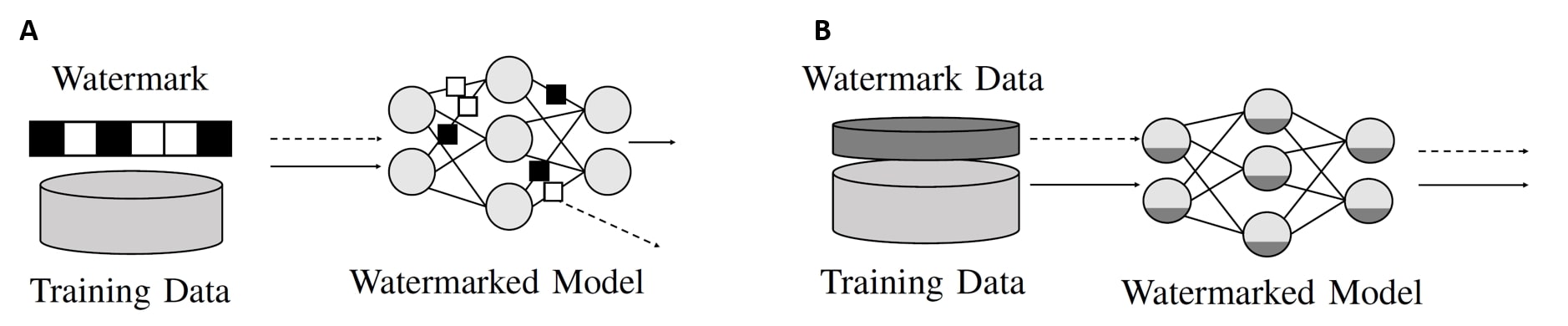}
\caption{Two broad approaches for watermarking ML models. (\textbf{A}) Define a watermarking bit string and embed it into the model parameters. For verification retrieve bit values from parameters and compare result with original string. (\textbf{B}) Train the model on the original data and a separate watermarking trigger dataset. For verification query the trigger dataset and verify the labels w.r.t. the trigger dataset labels.}
\label{fig:2approaches}
\end{figure*}

To determine the different classes, the following five dimensions are considered: 
\begin{enumerate}
    \item \emph{Embedding method}: Refers to the method used to include the watermark into the model. 
    \item \emph{Verification access}: Specifies how the watermark can be verified, either through white-box or black-box access.
    \item \emph{Capacity}: Distinguishes between \emph{zero-bit} and \emph{multi-bit} schemes. The former refers to watermarks that do not carry additional information whereas the latter do.
    \item \emph{Authentication}: Indicates if the watermark directly allows the legitimate owner to prove ownership.
    \item \emph{Uniqueness}: States if single (stolen) model instances should be uniquely identifiable. 
\end{enumerate}

In the following, the five dimensions are characterized more in detail.

\subsection{Embedding Method}
\label{sub:embedding}

\begin{figure}[t]
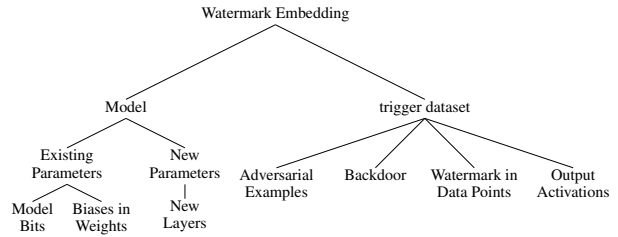

  \centering  
  \resizebox{0.45\textwidth}{!}{
  \Tree[.{Watermark Embedding} 
            [.Model 
                [.{Existing\\Parameters} 
                        [.{Model\\Bits} ]
                        [.{Biases in\\Weights} ]
                        ]
              [.{New\\Parameters} [.{New\\Layers} ]]]
          [.{trigger dataset} [.{Adversarial\\Examples} ]
                        [.{Backdoor} ]
                        [.{Watermark in\\Data Points} ]
                        [.{Output\\Activations} ]
                ]]
                }
  \caption{Taxonomy-tree depicting the methods that can be used to insert a watermark into an ML model.}
  \label{fig:methodtree}
\end{figure}

Watermarking techniques that have been proposed so far can be divided into two broad categories, namely~(1) inserting the watermark or related information directly into the model parameters, and~(2) creating a \emph{trigger}, \emph{carrier} or \emph{key} dataset, which consists of data points that evoke an unusual prediction behavior in the marked model.
See \cref{fig:2approaches} for a visualization of both concepts.
In~(1), the watermark might either be encoded in existing model parameters, for example, in the form of a bit string, or be inserted through adding additional parameters that rely on, or directly contain the watermark.
For (2), the trigger dataset needs to be fed along the original training data during the training process of the model.
Thereby, a backdoor is inserted into the model, such that the model learns to exhibit an unusual prediction behavior on data points from the trigger dataset. 
The unusual behavior can then be used in order to identify illegitimate model copies.
Therefore, when testing a model, the legitimate owner can query the trigger dataset and calculate the percentage of agreement between the model's prediction on the trigger dataset and the original corresponding labels.
If the resulting percentage exceeds a certain threshold (should be close to $1$), then the model is likely to be an illegitimate copy~\citep{Yang.2019Effectiveness}.
The trigger dataset can be generated independently, or based on the original training data.
Hence, it can potentially belong to a different data distribution than the training data.
Some watermarking schemes, for example \citep{Fan.2019Rethinking}, also combine both embedding categories.
Figure~\ref{fig:methodtree} shows a taxonomy-tree depicting a more fine-grained division of sub-categories within~(1) and~(2).
This taxonomy-tree provides the structure for presenting existing watermarking schemes more in detail in \cref{sec:techniques}.
In addition to the two broad categories of embedding watermarks into NNs, it is also possible to use  existing features of the models themselves as, so-called \emph{fingerprints}, to identify potentially stolen model instances \citep{Lukas.2019Deep,Chen.2019DeepMarks}.
Since these methods do not require explicitly inserting additional information as watermarks into the models, they will only be mentioned briefly in this document.

\subsection{Verification Access}
The type of access to a model required in order to perform watermark verification is closely related with the embedding method that was used to insert the watermark into it.
There exist two broad scenarios for watermark verification, namely \emph{white-box} and \emph{black-box}.
In a white-box scenario, a legitimate model owner needs  access to the model parameters in order to check for the watermark in potentially stolen copies of a model.
This might be necessary when the watermark is embedded into the model parameters alone and does not reflect in the model behavior.
However, in many scenarios, white-box access for verification is no realistic assumption.
A more realistic scenario is a black-box access scenario in which a legitimate model owner can access the potentially stolen model solely through a predefined query interface.
Through such an interface, the model owner could query (parts of) the watermark trigger and recognize a stolen instances of the model by its prediction behavior on these data points.
When selecting an adequate watermarking method, the access scenario for verification needs to be taken into account.
This is because a watermark that requires the model owner to have access to the model parameters for verification might be of little use in a scenario where the attacker deploys the stolen model in a black-box setting.

\subsection{Capacity}
Similar to \cite{Xue.2020DNN}, this work defines capacity as the watermark's capability to carry information. 
In general, a distinction can be made between \emph{zero-bit} and \emph{multi-bit} watermarking schemes.
Zero-bit watermarks do not carry additional information, such that they solely serve to indicate the presence or the absence of the watermark in a model.
An example for such a scheme could be using plain random data points as a trigger dataset to backdoor a model and verifying potentially stolen model instances by querying these data points and observing the model's predictions on them.
In multi-bit schemes, the watermark can carry information, \eg in the form of a bit string.
Thereby, such schemes can be used, among others, for creating a link between a model owner's identity and the watermark, or to mark individual model instances.

\subsection{Authentication}
By creating a link between a model owner's identity and the watermark, a watermarking scheme can serve to authenticate the legitimate owner.
This allows to extend the information \emph{that} a model was watermarked by the information \emph{by whom} it was watermarked which can be useful, for example, if the legitimate model owner wants to claim copyright in front of a legal entity.
The link between the owner's identity and the watermark can be expressed, among others, by including the owner's digital signature directly in the watermark or in the trigger data.
Besides enabling the legitimate owner to proof their ownership, watermarking schemes that offer authentication also prevent attackers from claiming ownership of existing watermarks, \ie \emph{forging} them.
Preventing forging is necessary to guarantee unambiguous ownership claims, see \Cref{sub:forging}. 
Watermarking schemes that do not inherently allow for authentication need to take different measures to prevent attackers from forging watermarks. 

\subsection{Uniqueness}
The last dimension along which to classify watermarking schemes for NNs concerns the uniqueness of the watermarks, \ie the question whether all instances of a model use the same watermark or if every instance receives a unique identification.
A crucial shortcoming of the former is that when a stolen copy of a model appears somewhere, it is impossible for the legitimate model owner to identify which of the parties that had access to the model stole it.
Unique watermarks allow to distinguish between different model instances, and thereby, enable a more fine-grained tracking of the intellectual property.
The distinction between different instances of a model can be, for example, implemented through the use of unique model identifiers or serial numbers \citep{Xu.2019novel}.
The watermark, then, does not only signal to the legitimate owner \emph{that} the model was stolen, but also \emph{by whom} it was stolen.

\section{Threat Model and Security Goal}
\label{sec:threat}
In general, as well as for watermarking schemes, the security of a system should be evaluated with respect to a specific threat space that characterizes the attacker's knowledge, capabilities, and objectives, and the underlying security goals.
This serves to thoroughly understand what properties the watermarking scheme needs to exhibit in order to adequately serve the goal of protecting a model in a given scenario:
For example, a watermarking scheme supposed to protect an ML model that is directly distributed to the users in a white-box fashion will most likely need to possess different properties than a scheme applied to a model that is only accessible in a black-box fashion.
This section, therefore, presents a unified threat model for watermarking schemes by elaborating the concrete requirements for watermarking, and properly characterizing the attack surface and the attacker.

\begin{table*}[ht]
  \caption{Requirements for watermarking techniques. Definitions adopted from \cite{Uchida.2017Embedding,Adi.2018Turning,Chen.2019DeepMarks,Li.2019Piracy}.}
  \label{tab:watermarkingRequirements}
  \centering
  {\renewcommand{\arraystretch}{1.5}
\begin{tabularx}{\textwidth}{lXX}
    \toprule
    \textbf{Requirement} & \textbf{Explanation} & \textbf{Motivation} \\
    \midrule
    Fidelity & Prediction quality of the model on its original task should not be degraded significantly & Ensures the model's performance on the original task \\  
    \rowcolor{Gray}
    Robustness & Watermark should be robust against removal attacks & Prevents attacker from removing the watermark to avoid copyright claims of the original owner \\
    Reliability & Exhibit minimal false negative rate & Allows legitimate users to identify their intellectual property with a high probability\\
    \rowcolor{Gray}
    Integrity & Exhibit minimal false alarm rate & Avoids erroneously accusing honest parties with similar models of theft \\
    Capacity & Allow for inclusion of large amounts of information & Enables inclusion of potentially long watermarks \eg a signature of the legitimate model owner\\
    \rowcolor{Gray}
    Secrecy & Presence of the watermark should be secret, watermark should be undetectable & Prevents watermark detection by an unauthorized party\\
    Efficiency & Process of including and verifying a watermark to ML model should be fast & Does not add large overhead \\
    \rowcolor{Gray}
    Generality & Watermarking algorithm should be independent of the dataset and the ML algorithms used & Allows for broad use \\
    \bottomrule 
\end{tabularx}}
\end{table*}

\subsection{Requirements for Watermarking}

In watermarking, the security goals can be expressed in the form of concrete requirements for an effective watermarking scheme. 
Within the last years, several such requirements have been formulated by different parties, \eg \cite{Uchida.2017Embedding,Adi.2018Turning,Chen.2019DeepMarks,Li.2019Piracy}.
See \cref{tab:watermarkingRequirements} for a structured overview on the requirements and their practical implications.
Note that it might not be feasible to implement all requirements simultaneously, since they might interfere with each other.
Take as an example \emph{reliability} and \emph{integrity}.
In order to make a watermark reliable, verification should be very sensitive and indicate ownership also in case of doubt. 
A watermarking scheme in which verification always and for every model under test indicates ownership would be perfectly reliable since it would detect every stolen model instance. 
However, that scheme would exhibit a very high false alarm rate, and erroneously accuse many honest parties of theft, which represents a violation of the integrity.

\subsection{Watermarking Attack Surface and Attacker}
The attack surface needs to be characterized to understand at what point and how an attacker might attempt to bypass the watermark.
Independent of the attacker, the main aspect to be considered is the scenario in which the model is stolen, \ie black-box or white-box access as depicted in  Figure~\ref{fig:access}.
A white-box scenario holds the advantage that by staling the model as is, the attacker is likely to retain the watermark within it.
This might render watermark verification for the legitimate owner more successful if the attacker does not employ additional methods to impede verification.
In a black-box scenario, \eg through model extraction (see \Cref{sub:modelextraction} and Figure~\ref{fig:stealing}), there is not necessarily a guarantee that the watermark is entirely transferred to the surrogate model.
As a consequence, watermark accuracy might already be degraded in the extracted model without additional explicit measures by the attacker.
Hence, watermark verification might be more difficult for the original model owner.
Further aspect of the attack surface are shaped by the attackers, their knowledge, capabilities, and objectives.

\subsubsection{Attacker Knowledge}
\label{sub:attackerknowledge}
The attacker knowledge refers to the information that an attacker holds about the system.
In NN watermarking, the information can consist of the following (from weak to strong): (1) the \emph{existence of the watermark}, knowledge on the (2) \emph{model and its parameters}, (3) the \emph{watermarking scheme used}, (4) (parts of) the \emph{training data}, and (5) (parts of) the \emph{watermark itself} or the \emph{trigger dataset}.
More meaningful information can, potentially, allow for more effective attacks.
For example, the sheer knowledge of the existence of a watermark within a model, without further details, will hardly serve as a warning to the attacker to release the stolen model with care in order to avoid detection by the legitimate owner.
However, when knowing what watermarking scheme was employed, and even possessing additional training data or parts of the watermark itself, an attacker might be able to more successfully extract a model in a black-box scenario, and to apply concrete attacks against the watermark, such as the ones described in \Cref{sec:attacks}.

\subsubsection{Attacker Capabilities}
Other than the information on the system, also the attacker's capabilities can shape the threat space.
A typical distinction here is to be made between a \emph{passive} and an \emph{active} attacker.
A passive attacker can not interact directly with the target model but might be able observe the model's behavior through its outputs or inputs-output pairs.
Such an attack is usually referred to as \emph{eavesdropping} and might yield sub-optimal attack results, or provide limited information on the watermark.
An active attacker, in contrary, might be able to interact with the model by (1) posing queries and (2) observing the corresponding output.
Thereby, through carefully choosing the inputs, the attacker might gain much more information on the target model and the watermark than by simple eavesdropping.
Also among active attackers, there is a broad range of capabilities.
For (1), namely providing model inputs, the number of queries that the attacker can pose might be restricted or unrestricted. 
Being able to pose more queries might enable the attacker to gain more information on the system.
There might, furthermore, exist restrictions on the type of queries that an attacker can pose, such as on the input format, the input source, or the range of possible input values.
When there are no restrictions on the type of queries, the attacker can query any possible data point to the model.
For (2), namely the observation of the model output, the attacker might also have different capabilities ranging from, for example, the ability to only observe the predicted class of a classifier, up to obtaining much more fine-grained information from the model, such as confidence scores for all classes.
Depending on the level of details that the model returns together with the predictions, the attacker's ability to learn about the system might vary.

\subsubsection{Attacker Objectives}
The last part of the threat space specifies the attacker's objectives.
This includes the question what for and where the stolen model will be used.
If the attacker plans to deploy it secretly and with no external interaction, there is a large chance that the model theft might remain uncovered.
Otherwise, if access to the stolen model is offered, for example, if the attacker wants to sell services that are based on the model and its predictions, the legitimate owner is more likely to successfully re-identify the stolen model instance.
In case a stolen model is exposed for interaction, the attacker's methods to prevent the model owner from successfully verifying the watermark play a vital role.
They are depicted more in detail in the following section.

\section{Attacks against Watermarking}
\label{sec:attacks}

\begin{figure*}[ht]
  \centering
  \includegraphics[width=.8\textwidth]{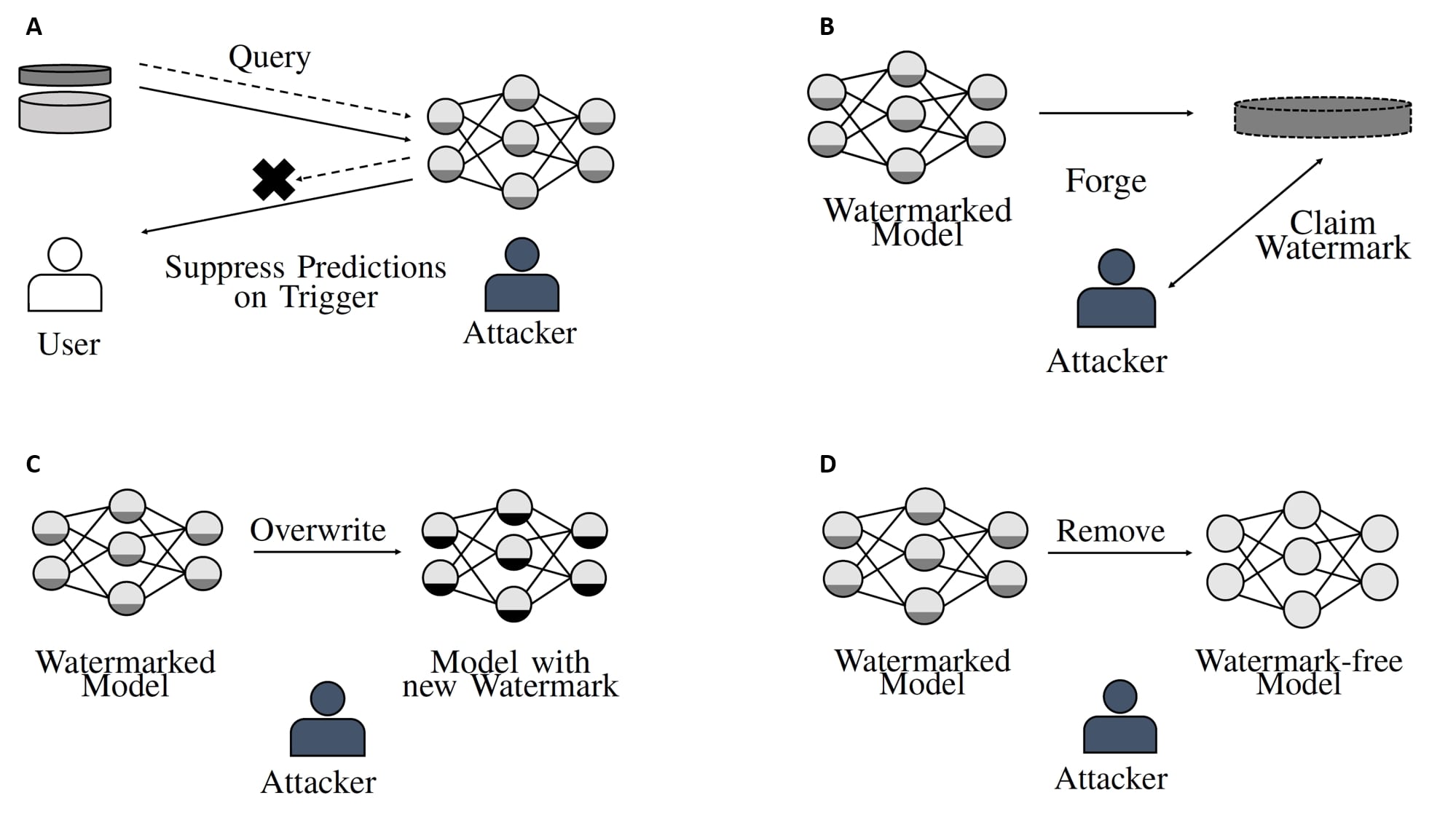}
\caption{Methods an attacker can apply to prevent the detection of a watermark in a stolen model. (\textbf{A}) Suppress watermark. (\textbf{B}) Forge watermark. (\textbf{C}) Overwrite watermark. (\textbf{D}) Remove watermark.}
\label{fig:preventdetection}
\end{figure*}

The attacks described in this section highlight several practical considerations that must be taken into account when designing watermarking methods that should still enable identification of stolen model instances, even when an attacker tries to prevent successful verification.
The attacks can be grouped into five different classes (from weak to strong), namely watermark \emph{detection}, \emph{suppression}, \emph{forging}, \emph{overwriting}, and \emph{removal}.
While watermark detection is a passive attack in the sense that it does not directly impede successful watermark verification, all other attacks actively try to reduce the watermark's suitability to prove the legitimate model owner's copyright claim.
Figure~\ref{fig:preventdetection} provides a visualization of the concepts behind the active attacks against watermarking schemes.

\subsection{Watermark Detection}
The weakest form of attack is concerned with the detection of watermarks in ML models. 
As stated above, watermark detection is a passive attack that does not directly prevent successful watermark verification by the legitimate model owner.
However, discovering the presence of a watermark in a stolen model increases the attacker's knowledge (see \Cref{sub:attackerknowledge}) and can, thus, be used as a base for further attacks.
Additionally, knowing that a stolen model was watermarked gives the attacker the opportunity to adapt this model's behavior in order avoid successful watermark verification.
To detect watermarks included in the target model's parameters, for example, \emph{property inference attacks} have shown to be successful \citep{Wang.2019Robust, Shafieinejad.2019On}, while backdoor detection can help to identify when models were watermarked by a trigger dataset \citep{aiken2021neural}.

\subsection{Watermark Suppression}
One way to avoid successful watermark verification can consist in suppression of the watermark in a stolen model instance. 
In a scenario where the legitimate model owner has white-box access to the model for watermark verification, the attacker, therefore, needs to dissimulate any presence of a watermark in the model parameters and behavior.
When the legitimate model owner has black-box access to the potentially stolen model, suppressing the reactions of the model to the original watermark trigger might be sufficient for an attacker to prevent detection.
This can be achieved, for example, by identifying possible trigger data points and modifying the model's predictions on them \citep{Namba.2019Robust, Hitaj.2018Have}.
Additionally, so-called \emph{ensemble attacks} that rely on stealing $n$ models from different providers and using them as an ensemble for prediction, have been shown successful for watermark suppression, since they eliminate individual watermark triggers  \citep{Hitaj.2018Have}.

\subsection{Watermark Forging}
\label{sub:forging}
In some cases, the attacker might also be able to forge a watermark on a given model. 
This attack does not necessarily prevent the legitimate model owner from successfully verifying the watermark in a stolen model instance.
However, it creates an ambiguity, in which, for an external entity, such as a legal authority, it is not possible anymore to decide which party has watermarked the given model.
Thereby, the attack prevents the legitimate owner from successfully claiming copyright of the intellectual property.
Watermark forging might be done by (1) recovering the legitimate owner's watermark and claiming ownership (if there is no binding between the watermark and the owner) \citep{Xu.2019novel}, (2) adding a new watermark that creates ambiguity concerning ownership \citep{Fan.2019Rethinking}, or (3) identifying a fake watermark within the model that coincidentally acts like a real watermark but actually is not \citep{Guo.2018Watermarking}.

\subsection{Watermark Overwriting}
Within the attack setting of watermark overwriting, an attacker knows that the stolen model was watermarked, but might have no knowledge about the legitimate owner's concrete watermark or trigger dataset.
The attacker can then embed a new watermark into the model to pretend ownership.
Note that the new watermark does not necessarily need to be created with the same watermarking scheme as the original one.
In a weak form of the attack, the original watermark is still present after embedding the new one, such that both watermarks co-exist in the model.
Therefore, the legitimate owner can then no longer prove (unique) ownership due to ambiguity.
In a stronger version of this attack, the attacker might be able to replace the original watermark entirely with the new one \citep{Wang.2018Attacks}.
Thereby, the ownership claim can be taken over completely \citep{Li.2019Piracy}.

\subsection{Watermark Removal}
\label{ssec:removal}
As an ultimate solution to prevent successful watermark verification by a legitimate model owner, an attacker might also try to entirely remove the watermark from a stolen model \citep{Zhang.2018Protecting}.
The success of the removal attack usually depends on the attacker's knowledge about \emph{(i) the presence of a watermark}, \emph{(ii) the underlying watermarking scheme}, and on the \emph{(iii) availability of additional data, \eg for fine-tuning or retraining}.
Especially the last point is interesting to consider because many attacks presented in literature rely on the assumption that an attacker has large amounts of data available to fine-tune \citep{Chen.2019REFIT}.
In reality, an attacker possessing a sufficiently large dataset to train a good model might be less motivated to steal a model, instead of training it from scratch~\citep{Chen.2019REFIT}.

The most general methods that can be used to remove watermarks can be grouped as follows:
\begin{itemize}
  \item \emph{Fine-Tuning:} use initial model parameters and fine-tune them to a refinement set.
  Thereby, it is possible to improve a model for certain kinds of data \citep{sharif2014cnn, tajbakhsh2016convolutional}.
  This process might remove the watermark when model parameters or prediction behaviors are changed \citep{Chen.2019REFIT, Shafieinejad.2019On}.
  \item \emph{Pruning \citep{augasta2013pruning, molchanov2019pruning}:} cut some redundant parameters and obtain a new model that looks different from the original model but still has a similar high prediction accuracy.
  If the parameters containing the watermark are cut, it is no longer possible to verify the watermark  \citep {Zhang.2018Protecting}.
  \item \emph{Rounding \citep{guo2018survey}:} reduce the precision of the parameters.
  If the model strongly overfits the watermark triggers, or the watermark is included in the parameters directly, rounding might destroy the watermark information \citep{Yang.2019Effectiveness}.
  \item \emph{Fine-Pruning:} first prune the model architecture, then continue to train. 
  In the benign setting, this helps recover some of the accuracy that may have been lost during pruning. 
  In the presence of backdoors, such as certain watermarks, this also contributes to overwrite their information \citep{liu2018fine, Jia.2020Entangled}.
  \item \emph{Model Compression \citep{chen2015compressing,han2016eie}:} optimize the memory needed to fit the model while preserving its accuracy on the task. 
  This might be necessary in mobile or embedded devices with limited resources.
  Model compression is performed by, for example, removing insignificant parameters and pruning links between neurons.
  This can affect the watermark reliability, especially if the neurons used for the watermarking task are different from the ones of the original task, because then, they can be pruned without losing accuracy in the original task \citep{Yang.2019Effectiveness}.
  \item \emph{Distillation \citep{Hinton.2015Distilling}:} transfer the prediction power of a potentially very complex teacher model to a less complex student model.
  This approach finds application, for example, in low-power environments, where simpler models are to be preferred.
  It can, however, not be guaranteed that the watermark is also included in the simpler model \citep{Yang.2019Effectiveness}.
  \item \emph{Transfer Learning \citep{oquab2014learning}:} update the classification task of a model to a related but slightly different task.
  Therefore, model layers towards the output are modified.
  This approach saves computational power because large parts of trained models' weights can be applied to the new task with solely small changes.
  However, the changes within the model layers can lead to a removal of the watermark \citep{Adi.2018Turning}.
  \item \emph{Computation Optimization \citep{jaderberg2014speeding}:} reduce computation time of NNs, for example, by low-rank expansion techniques to approximate convolution layers. 
  The reduction might as well lead to a watermark removal \citep{Yang.2019Effectiveness}.
  \item \emph{Backdoor Removal \citep{liu2021removing}:} remove backdoors, \ie functionalities in the NN that are not relevant for the original task (see Section~\ref{sub:backdoors}).
   \cite{Li.2019Piracy} pointed out that if the watermarking task is indeed a backdoor function that is too loosely related with the original task, it is possible to remove the watermark by normal backdoor removal attacks against NNs, such as \cite{Wang.2019Neural}.
  \item \emph{Retraining:} an ML model might be trained continuously, instead of being trained once and then released for prediction.
  Through retraining, models can adapt to potential shifts in the underlying data distribution over time.
  While retraining, the watermark might be damaged  \citep{LeMerrer.2019Adversarial, Rouhani.2018DeepSigns, Adi.2018Turning, Zhang.2018Protecting}.
\end{itemize}

Additionally to the above-mentioned approaches, there also exist more specific attacks proposed in literature that rely, for example, on regularization \citep{Shafieinejad.2019On}, or on graph algorithms \citep{Wang.2020Watermarking}.

\section{Categorizing Watermarking Methods}
\label{sec:techniques}

This section surveys examples of watermarking methods proposed in the literature to illustrate and validate the taxonomy presented in Section~\ref{sec:taxonomy}.
The methods are presented in semantic groups based on their embedding method and their distinctive characteristics.
Methods that might fit to several groups are presented according to their most distinctive property.
For a more comprehensive overview of existing approaches, see~\cite{Li.2021survey}.

\subsection{Embedding Watermarks into Model Parameters}

Early approaches to mark ML models with the aim of including information about the training data into the model parameters were proposed by \cite{Song.2017Machine}.
Among others, they showed how to include information in the least significant bits of the model parameters or the parameters' signs, and developed a \emph{correlated value encoding}, to maximize a correlation between the model parameters and a given secret.  
A similar method was then applied by  \cite{Uchida.2017Embedding} as the first explicit watermarking scheme in NNs.
The authors interpret the watermark as a $T$-bit string $\{0,1\}^T$.
In order to include it into the model parameters, they use a composed loss function $L(\theta) = L_{O} + \lambda L_R(\theta)$ with $L_{O}$ being the loss of the original task and $L_R$ an \emph{embedding regularizer}.
This regularizer imposes a statistical bias on certain model parameters in order to represent the watermark.
To project the weights carrying the watermark information, an \emph{embedding parameter} $X$ is used as a secret key needed for watermark embedding and verification. 

\begin{table*}[ht]
  \caption{Techniques to \emph{embed watermarks into model parameters} sorted alphabetically by author.}
  \label{tab:watermarkParameters}
  \centering
  {\renewcommand{\arraystretch}{1.5}
\begin{tabularx}{\textwidth}{Xlllcc}
    \toprule
    \textbf{Method} 
    &  \textbf{Verification} &  \textbf{Method} &  \textbf{Capacity} & \textbf{Auth.} &  \textbf{Unique}\\
    \midrule 
     \cite{Uchida.2017Embedding}: Embed bit string watermark to random model parameters' statistical bias  
    & White-box & Biases in Weights & Multi-Bit & No & No \\
    \rowcolor{Gray}
  \cite{Fan.2019Rethinking}: Adding passport layers into NNs 
    & W \& B-box  & New Layers & Multi-Bit & Yes & No\\   
  \cite{Song.2017Machine}: Include information in model parameters  \eg least significant bit or sign) 
    & White-box  & Model Bits & Zero-Bit & No & No\\
    \rowcolor{Gray} 
  \cite{Wang.2019Robust}: Create non-detectable watermarked parameters   
    & White-box & Existing Parameters & Zero-Bit & No & No\\ 
  \cite{Wang.2020Watermarking}: Extend \cite{Uchida.2017Embedding}, include watermarks into quickly converging model parameters   
    & White-box & Existing Parameters & Zero-Bit & No & No\\ 
    \bottomrule 
\end{tabularx}}
\end{table*}

 \cite{Wang.2020Watermarking} extend this work by developing an alternative for the embedding parameter $X$.
Instead, they employ an additional independent NN on selective parameters from the original model to project the watermark.
For training of the original model, they use the above-mentioned loss function $L(\theta) = L_{O} + \lambda L_R(\theta)$.
To train the additional NN, they apply the binary cross-entropy loss between its output vector and the watermark.
The additional NN is not released publicly and serves for watermark verification.
\cite{Wang.2018Attacks}, however, showed that both the approaches of \cite{Uchida.2017Embedding} and \cite{Wang.2020Watermarking} do not meet the requirement of watermark secrecy because they cause easily detectable changes in the statistical distribution of the model parameters.

 \cite{Wang.2019Robust} proposed a strategy to create undetectable watermarks in a white-box setting based on generative adversarial networks (GAN).
The watermarked model $h_{O}$ serves as the generator, whereas a watermark detector that detects changes in the statistical distribution of the model parameters serves as a discriminator $h_{D}$.
During training, $h_{O}$ is encouraged to generate non-detectable watermarks whereas $h_{D}$ tries to distinguish watermarked from non-watermarked models.
Both optimize the following functions respectively:
\begin{align}
h_{O}: min(L_{O}(\theta) + \lambda L_R(\theta) -\lambda_2 \log h_{D}(\theta)) \\
h_{D}: max(\log h_{D}(\theta_{non}) + \log(1- h_{D}(\theta)))
\end{align} 
where $\theta_{non}$ refers to the parameters of non-watermarked previously trained models.

\cite{Fan.2019Rethinking} suggested embedding \emph{passport-layers} with digital signatures into NNs for ownership verification.
The passport layers are inserted into convolutional neural networks (CNNs) 
and calculate hidden parameters without which the model's inference accuracy is reduced.
For verification, the authors developed three mechanisms based on different strategies of distributing and verifying the passports.

\cref{tab:watermarkParameters} provides an overview on the mentioned methods that rely on inserting watermarks directly into the model parameters.

\subsection{Using Pre-Defined Inputs as Triggers}

\begin{figure}[t]
  \centering
  \includegraphics[width=.45\textwidth]{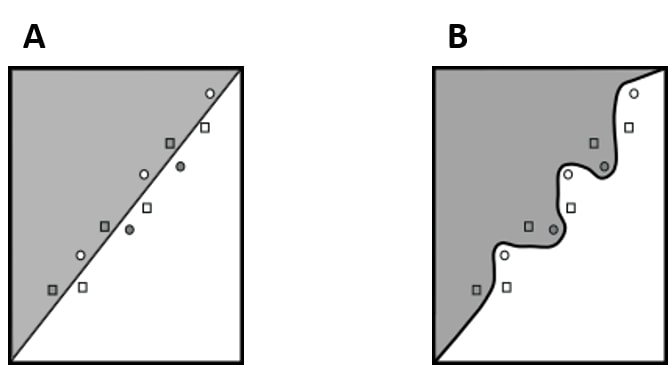}
\caption{Visualization of  \citep{LeMerrer.2019Adversarial}, figure adapted from \cite{LeMerrer.2019Adversarial}. Squares indicate adversarial examples for the corresponding color, circles correspond data points that lie close to the decision boundary but are correctly classified. Decision boundary is altered to correctly classify the adversarial examples. (\textbf{A}) Original decision boundary. (\textbf{B}) After fine-tuning.}
\label{fig:leMerrer}
\end{figure}

 \cite{LeMerrer.2019Adversarial} proposed directly marking the model's action itself by slightly moving the decision boundary through adversarial retraining such that specific queries can exploit it.
Therefore, their approach first identifies adversarial samples and normal data points that are very close to the decision boundary. 
Then, the trigger dataset is composed by $50\%$ of the adversarial examples, and $50\%$ of the data points that do not cause misclassification, but are close to decision boundary.
Afterwards, the trained classifier is fine-tuned to predict the trigger data points to their correct original class.
See \cref{fig:leMerrer} for a visualization of this approach.
The resulting labeled data points are supposed to serve as an expressive trigger dataset.
 \cite{Namba.2019Robust} argue that this method offers weak integrity because, nowadays, adversarial retraining is broadly used to create more robust models, hence a non-watermarked model can be mistaken to be watermarked.

 \cite{Adi.2018Turning} consider watermarking from a cryptographic point of view.
The authors generate abstract color images with randomly assigned classes as a trigger dataset.
In order to guarantee non-trivial ownership, a set of commitments is created over the image/label pairs before embedding the watermark into the model.
Thereby, at verification time, ownership can be proven by selectively revealing these commitments.
A similar approach for creating the watermark trigger dataset was also described by \cite{Zhang.2018Protecting}.
They include irrelevant data samples, \eg from another unrelated dataset, as watermarks into the training data.
Those samples are labeled with classes from the original model output.
During training, the model learns to assign real images and those trigger samples to the corresponding classes.

 \cite{Rouhani.2018DeepSigns} developed an approach of including the watermark as a $T$-bit string into the probability density function (\emph{pdf}) of the data abstraction obtained in different network layers.
These layers' activation maps at intermediate network layers roughly follow Gaussian distributions.
The legitimate model's owner can choose in how many of those they want to embed the watermark string.
Afterwards, the network is trained to incorporate the watermark information in the mean values of the selected distributions.
A projection matrix $A$ can be used to map the selected distribution centers to the binary watermark vector.
In a white-box setting, this matrix $A$ is utilized for verification.
For black-box verification, a trigger dataset can be constructed from data points whose features lie in the model's unused regions, \ie samples at the tail regions of the pdf.
In contrast to methods including the watermark in the static model content, like  \cite{Uchida.2017Embedding}, this approach changes the dynamic content of the model, namely the activations that depend on the data and the model.
This results in a more flexible and not that easily detectable change \citep{Rouhani.2018DeepSigns}. 

\cite{Chen.2019BlackMarks} proposed taking the model owner's binary signature as a watermark for an NN.
Their aim is to combine the advantage from black-box and white-box watermark extraction, \ie weaker assumptions on the attacker's power and large capacity at the same time.
To include the signature into the model, they build a model-dependent encoding scheme that clusters the model's output activations into two groups according to their similarities, one group for class~$0$, and one for class~$1$.
The binary signature is then included in the model's output activations and can be verified through a designated trigger dataset that can be passed to the model in order to retrieve the signature.

\subsubsection{Trigger Dataset Creation Based on Original Training Data}

\begin{table*}[ht]
  \caption{Techniques to embed \emph{watermarks into the training data} in order to create the trigger dataset.}
  \label{tab:watermarkData}
  \centering
  {\renewcommand{\arraystretch}{1.5}
\begin{tabularx}{\textwidth}{Xlllcc}
    \toprule
    \textbf{Approach} 
    &  \textbf{Verification} &  \textbf{Method} &  \textbf{Capacity} & \textbf{Auth.} &  \textbf{Unique}\\
    \midrule 
     \cite{Guo.2018Watermarking}: Model owner's signature embedded into the training data 
    & Black-box & Watermark in Data Points & Multi-Bit & Yes & No\\
    \rowcolor{Gray}
     \cite{Li.2019How}: Include ``logo'' in the training data images, use an auto-encoder to make trigger samples close to original data 
    & Black-box & Watermark in Data Points & Zero-Bit & Yes & No\\
     \cite{Zhang.2018Protecting}: Include (meaningful) information in the training data samples 
    & Black-box & Watermark in Data Points & Multi-Bit & Yes & No\\
    \bottomrule 
\end{tabularx}}
\end{table*}

Some watermarking approaches rely on inserting forms of digital media watermarks into the original training data in order to create the model's trigger dataset.
The approach by \cite{Guo.2018Watermarking} generates an $n$-bit signature of the model owner and embeds it into the training data in order to generate the trigger dataset.
The authors make sure that the altered images from the trigger dataset obtain different labels than the original data points that they are based on.

 \cite{Zhang.2018Protecting} described algorithms for watermarking NNs for image classification with remote black-box verification mechanisms.
One of their algorithms embeds meaningful content together with the original training data as a watermark.
An example for this approach is embedding a specific string (which could be the company name) into a picture of the training set when predicting images, and assigning a different label than the original one to the modified sample.
Instead of a meaningful string, it is also possible to embed noise into the original training data.
A similar approach to the first algorithm of  \cite{Zhang.2018Protecting} was proposed by \cite{Li.2019How} who combine some ordinary data samples with an exclusive `logo' and train the model to predict them into a specific label.
To keep these trigger samples as close as possible to the original samples, an autoencoder is used whose discriminator is trained to distinguish between training and trigger samples with the watermarks.
 \cite{Sakazawa.2019Visual} proposed a cumulative and visual decoding of watermarks in NNs, such that pattern embedded into the training data can become visual for an authentication by a third party.

See \cref{tab:watermarkData} for an overview on methods that use original training data to generate the trigger dataset.

\subsubsection{Robust Watermarking}
A problem of watermarking methods that rely on using a  trigger dataset from a different distribution than the original training data is that the models are actually trained for two different (and independent) tasks.
Research has shown that when these tasks are more or less unrelated, it is possible to remove the watermarks through attacks described in Section~\ref{ssec:removal} without affecting the model's accuracy on the original task learned through the training data \citep{Li.2019Piracy,Jia.2020Entangled}.

For example, \cite{Yang.2019Effectiveness} showed that distillation \citep{Papernot.2016Distillation} is effective to remove watermarks.
This results from the fact that the information learned from the watermark trigger dataset is redundant and independent of the main task.
Hence, this information is not transferred to the resulting surrogate model when doing distillation.
As a solution, the authors described an `ingrain'-watermarking method that regularizes the original NN with an additional NN that they refer to as ingrainer model~$g_\omega$ which contains the watermark information.
Regularization is performed with a specific ingrain loss 
$C(h_{\theta,T}(x),g_\omega(x))$ ($T$ is the temperature in the softmax) that causes the watermark information to be embedded into the same model parameters as the main classification task.
The joint loss function over the training data $D_s$ is given by:
\begin{align}
L_{D_s}(h_\theta) = \frac{1}{|D_s|}\sum_{x \in D_s} C(h_\theta(x),y)+\delta C(h_{\theta,T}(x),g_\omega(x))\text{,}
\end{align}
where the labels are indicated by $y$, and $\lambda$ determines the degree of ingrain.

\begin{table*}[t]
  \caption{Techniques using a \emph{specific trigger dataset} as a watermark sorted alphabetically by author.}
  \label{tab:watermarkTrigger}
  \centering
  {\renewcommand{\arraystretch}{1.5}
\begin{tabularx}{\textwidth}{Xlllcc}
    \toprule
    \textbf{Approach} 
    &  \textbf{Verification} &  \textbf{Method} &  \textbf{Capacity} & \textbf{Auth.} &  \textbf{Unique}\\
    \midrule 
     \cite{Adi.2018Turning}: Abstract color images with random classes as trigger dataset 
    & Black-box & Backdoor & Zero-Bit & Yes & No\\
    \rowcolor{Gray}
     \cite{Chen.2019BlackMarks}: Include model owner's binary signature in output activations 
    & Black-box & Output Activations & Multi-Bit & Yes & No\\
     \cite{Jia.2020Entangled}: Entangled watermarks through training with \emph{soft nearest neighbor loss}
    & Black-box & Backdoor & Zero-Bit & No & No \\
    \rowcolor{Gray}
     \cite{LeMerrer.2019Adversarial}: Adversarial decision boundary modification through trigger sample consisting of adversarial examples 
    &  Black-box & Adversarial Examples & Zero-Bit & No & No\\
     \cite{Li.2019Piracy}: Null embedding watermark consisting of a pixel pattern 
    & Black-box & Watermark in Data Points & Multi-Bit & Yes & No\\
    \rowcolor{Gray}
    \rowcolor{Gray}   
     \cite{Namba.2019Robust}: Exponential weighting: enforce watermark predictions with higher weights during training 
    & Black-box & Backdoor & Zero-Bit & No & No\\
     \cite{Rouhani.2018DeepSigns}: Include watermark in probability density function of network layers 
    & W \& B-box & Biases in Weights & Multi-Bit & No & No\\
    \rowcolor{Gray}
     \cite{Yang.2019Effectiveness}: Train additional model (called ingrainer) that contains the watermark information via its prediction on training data  
    & Black-box & Backdoor & Zero-Bit & No & No\\
    \bottomrule 
\end{tabularx}}
\end{table*}

 \cite{Jia.2020Entangled} proposed a similar idea that relies on `entangled watermarking embeddings'.
The entanglement is used to make the model extract common features of the data that represent the original task and the data that encodes the watermarks and stems from a different distribution.
Therefore, the authors apply the \emph{soft nearest neighbor loss} (SNNL) \citep{Salakhutdinov.2007Learning}.
Informally spoken, the SNNL measures \emph{entanglement} over labeled data, \ie how close pairs of points from the same class are relative to pairs of points from different classes \citep{Frosst.2019Analyzing}.
Points from different groups that are closer relative to the average distance between two points are called \emph{entangled}.
Using entanglement when including a watermark ensures that the watermark and the original task are represented by the same sub-model and not by different ones that may be harmed during extraction.
Hence, it becomes more difficult for an attacker to extract the model without its watermarks.
At the same time, through the entanglement, removing the watermark would result in a decrease of model performance on the original task.
 
 \cite{Namba.2019Robust} described a method they called `exponential weighting'.
They generate a watermark trigger by random sampling from the training distribution and assigning wrong labels to that sample for training. 
To protect the watermark against pruning or retraining attacks, the authors propose embedding the samples by exponential weighting, \ie imprint trigger samples with greater force and cause the model to learn them profoundly.
Therefore, they increase the weight of the model parameters that are involved in the prediction exponentially, and thereby make the prediction depend mainly on some few and very large model parameters which are harder to change through the mentioned attacks.

\begin{figure}[t]
  \centering
  \includegraphics[width=.15\textwidth]{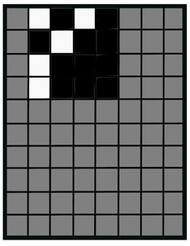}  
  \caption{Watermark pattern, figure adapted from \cite{Li.2019Piracy}.}
  \label{fig:Li}
\end{figure}

\cite{Li.2019Piracy} developed a `null embedding' for including watermarks into the model's initial training, such that attackers are not able to remove them or include their own watermarks on top. 
Therefore, they generate a filter pattern $p$ as shown in \cref{fig:Li}.
Image pixels under the white pattern pixels are changed to a very large negative number, image pixels under black pattern pixels are changed to a very large positive number, and pixels under gray pattern pixels stay unchanged.
Over this process, the predicted class of the image needs to stay the same as for the original image.
Using extreme values and setting strong deterministic constraint on the optimization during learning leads to strong watermark inclusion.
To create a binding between the owner and the pixel pattern, the authors propose using the owner's signature and a deterministic hash function to generate the pattern \citep{Li.2019Piracy}.

See \cref{tab:watermarkTrigger} for an overview on all the mentioned methods that rely on using a trigger dataset to watermark ML models.

\begin{table*}[ht]
  \caption{Techniques to generate unique watermarks that can be verified by querying a trigger dataset.}
  \label{tab:watermarkUnique}
  \centering
  {\renewcommand{\arraystretch}{1.5}
\begin{tabularx}{\textwidth}{Xlllcc}
    \toprule
     \textbf{Approach} 
    &  \textbf{Verification} &  \textbf{Method} &  \textbf{Capacity} & \textbf{Auth.} &  \textbf{Unique}\\
    \midrule
     \cite{Chen.2019DeepMarks}: End-to-end unique watermarking scheme based on anti-collusion codebooks for individual users 
    & White-box & Biases in Weights & Multi-Bit & Yes & Yes \\
    \rowcolor{Gray}
     \cite{Xu.2019novel}: Embed unique serial number based on model owner's signature and create endorsement by certification authority 
    & Black-box & Backdoor & Multi-Bit & Yes & Yes\\
    \bottomrule 
\end{tabularx}}
\end{table*}

\subsubsection{Unique Watermarking}

The requirements for unique watermarking schemes are the same as the ones for common watermarking methods (see \cref{tab:watermarkingRequirements}), but extended by the two following points~\cite{Chen.2019DeepMarks}:
\begin{itemize}
    \item \emph{Uniqueness}. Watermarks should be unique for each user. 
    This serves to identify model instances individually.
    \item \emph{Scalability}. Unique watermarking schemes should scale to many users in a system.  
    This allows for a large-scale distribution of the target model.
\end{itemize}

 \cite{Chen.2019DeepMarks} proposed an end-to-end collusion-secure watermarking framework for white-box settings.
Their approach is based on anti-collusion codebooks for individual users which are incorporated in the pdf of the model weights.
The incorporation is achieved by using a watermark-specific regularization loss during training.

 \cite{Xu.2019novel} embed a serial number to NNs for model ownership identification.
Their solution generates a unique serial number as a watermark and creates an endorsement by a certification authority on it.
Serial numbers are generated by the owner through a digital signature algorithm (based on the owner's private key).
During model training, the serial number is fitted into the model, together with the original task by having a second loss, such that owner verification can be achieved by sending trigger inputs, extracting the serial number and verifying it with the certification authority.

See \cref{tab:watermarkUnique} for an overview on these methods generating unique watermarks.

\subsection{Using Model Fingerprints to Identify Potentially Stolen Instances}
\label{sub:fingerprints}
Instead of explicitly adding watermark information into an ML model, some methods use existing features of the model in order to identify potentially stolen instances.
This offers the advantages that no overhead is added to the original training task and that the model's original prediction abilities are not affected.
However, as those methods do not actively alter the model in order to include a watermark, they will only be mentioned very briefly in this document.

 \cite{Zhao.2020AFA} use adversarial examples as fingerprints for NNs.
They identify some special adversarial examples within NNs that they call `adversarial marks'.
Those adversarial marks differ from traditional adversarial examples in their transferability: they show high transferability between models that are similar, and low transferability between models that are different.
The authors argue that adversarial marks represent  suitable model fingerprints as they are difficult to remove due to the number and type of adversarial examples being practically infinite.

 \cite{Lukas.2019Deep} also exploit the transferability of adversarial examples in order to verify the ownership of ML models.
They define the class of \emph{conferrable} adversarial examples. 
Those examples transfer only to surrogate models of a target model (potential illegitimate copies), but not to reference models, \ie similar models trained on similar data for a related task.
By querying those examples to a model, one can identify whether this model is a copy of the target model or not.
The authors also propose a generation method for this class of adversarial examples and prove that this watermarking method is robust against distillation \citep{Papernot.2016Distillation} and weaker attacks.

\section{Discussion}
\label{sec:discussion}
This section, first, discusses the pros and cons of existing classes of watermarking methods.
Afterwards, it revisits the requirements for effective watermarking to provide a structured reasoning for choosing or designing adequate watermarking schemes.
Finally, it presents limitations of existing methods and proposes an outlook on promising research directions.

\subsection{Discussing Potential Shortcomings}

First, in trigger dataset-based watermarking approaches, watermark detection relies on the model's reaction on queries from the trigger dataset.
If the agreement of the prediction on them to the trigger dataset's original labels overpasses a given threshold, this suggests the presence of the watermark.
However, defining a suitable threshold to identify a stolen model requires thorough tuning.
If the threshold is set too high, slight modifications in a stolen model might already be sufficient to prevent watermark detection, which violates the reliability requirement.
If the threshold is set too low, different models might erroneously be identified as stolen model instances, which violates the integrity requirement.
Hence, the choice of a threshold also expresses a trade-off between reliability and integrity.

Additionally to the issue of choosing an adequate threshold, \cite{Hitaj.2018Have} formulated the general disadvantage for the scenario of public verification.
They argue that after the verification algorithm is run against a stolen model, the attacker is in possession of the trigger dataset, which enables to fine-tune the model on those data points to remove the watermark.
Hence, in order to run several verifications, the original trigger dataset needs to be divided, or there have to be several trigger datasets.
This class of approaches, thus, has its limitation due to the maximum amount of backdoors that can be embedded in an NN.
Approaches that need few queries, like \cite{Jia.2020Entangled}, might allow for a higher number of independent watermark verifications with the same model capacities.

Second, watermarking schemes embedding watermarks into the ML models' parameters without taking precautions do not only leave a detectable trace in the model and, hence, violate the secrecy requirement \citep{Wang.2018Attacks}, but they also often rely on white-box access for verification. 
Even though in some scenarios, the latter might be feasible, still, assuming black-box access often is a more realistic scenario.
Therefore, such schemes might be suitable only to very specific applications.

Third, watermarking schemes that do not exhibit a verifiable link between the watermark and the legitimate model owner enable an attacker to forge the watermark. 
A similar disadvantage exists for watermarking schemes that rely on non-specific data points as trigger data (\eg \citep{Adi.2018Turning,Guo.2018Watermarking}).
Those might enable an attacker to choose (random) different points than the initial watermark, in order to claim having marked the model with them as triggers.
Approaches that do not allow already marked models to be marked again, like \citep{Li.2019Piracy}, can prevent this threat.

Finally, due to their instability, their potentially low robustness against fine-tuning or retraining, and, in some cases, their transferability, that might violate watermark integrity, adversarial examples used for watermarking \citep{LeMerrer.2019Adversarial} or fingerprinting \citep{Zhao.2020AFA,Lukas.2019Deep} might exhibit important drawbacks.
 \cite{Namba.2019Robust} point out that especially an adaption of the model's decision boundary according to some adversarial examples, as in \cite{LeMerrer.2019Adversarial}, might be likely to violate the integrity requirement because its effect is similar to the effect of adversarial retraining, a method commonly used to make ML models more robust.

\subsection{Discussing Requirements}
Additionally to considering the pros and cons of existing classes of watermarking methods, this section discusses the question of choosing or creating reliable watermarking methods by revisiting the requirements presented in \cref{tab:watermarkingRequirements}.

\begin{itemize}
\item\emph{Fidelity:} To guarantee fidelity, existing watermarking schemes aim at preserving model performance on the original task.
Depending on the scheme, this can be achieved through different means, \eg only minimally altering the original decision boundary \citep{LeMerrer.2019Adversarial}, or including the watermark to early converging model weights \citep{Wang.2020Watermarking}.

\item\emph{Robustness:} 
If the trigger dataset stems from a significantly different distribution than the original data, as in \citep{Adi.2018Turning,Fan.2019Rethinking}, the model learns two different (and independent) tasks.
Therefore, it is possible to extract them independently, or to remove the watermark without causing an impact on the model's prediction performance.
Thus, to achieve robustness, watermarking schemes need to take measures to create a relation between both tasks, and to enforce the watermark to the model such that it cannot be removed easily.

\item\emph{Reliability:} Certain factors can influence watermark reliability.
First, similarly as for robustness, if the trigger dataset stems from a different distribution than the original dataset, reactions of the stolen model to the watermark triggers can be suppressed by an attacker.
Second, all schemes that rely solely on white-box verification, might offer lower reliability, as such access to all potentially stolen model instances might not always be a realistic assumption, which might prevent successful verification.

\item\emph{Integrity:} Quantifying watermarking schemes' integrity is a challenging task, as it requires judging how (potentially all other) non-watermarked models react on the given trigger dataset.
A good trigger dataset is characterized by the uniqueness of the watermarked model's predictions on it, in order to accuse no honest parties with similar models of theft.

\item\emph{Capacity:} Capacity expresses how much information can be included in the watermark. 
To allow for specific tasks, such as owner authentication, inserting multi-bit watermarks is common practice \citep{Guo.2018Watermarking, Zhang.2018Protecting}.

\item\emph{Secrecy:} Watermarking schemes that change the model parameters in a detectable way, as for example \citep{Uchida.2017Embedding}, violate the secrecy requirement.
To prevent watermark detection, adding them to the dynamic model content \citep{Rouhani.2018DeepSecure}, or taking measures to force the model parameters to stay roughly the same \citep{Li.2019How}, are possible solution.

\item\emph{Efficiency:} Efficiency can be evaluated with regard to embedding and verification time, \ie the overhead in training and the computation time, or number of queries needed to verify the watermark.
Most existing work does not explicitly evaluate computational overhead of their approaches.
 \cite{Jia.2020Entangled} present one of the few evaluations of efficiency, and come to the conclusion that, for their approach, the trigger dataset should consist of more than half the amount of data samples as the original data.
Therefore, the model needs to train with $150$-$200$\% of the original data. 
Especially for large datasets, this might result in large overhead.

\item\emph{Generality:} Not all existing schemes directly generalize to all datasets, for example \citep{Chen.2019BlackMarks} needs a different watermark encoding scheme on each dataset.
Such behavior can be considered as lacking generality. 
\end{itemize}

\subsection{Limitations and Outlook}
Based on the evaluation of existing schemes and their security requirements, this section presents current limitations and promising future research directions.

The largest limitation of watermarking in general is that it represents a passive defense.
\Ie, watermarking schemes cannot prevent theft, but only detect it afterwards. 
Some research was conducted in order to issue security warnings, once an ML model is about to reveal enough information that an attacker or a group of attackers might be able to extract its functionality \citep{Kesarwani.2018Model}.
Further research focused on creating models that solely achieve high accuracy when being queried by an authorized user \citep{Chen.2018Protect}.
Other work was directed towards the development of models that are more difficult to steal, \eg by only returning hard labels and no probabilities per output class, by perturbing the prediction outputs \citep{Orekondy.2019Prediction}, or by designing networks that are extremely sensitive to weight changes, which makes it difficult for an attacker to steal and adapt them \citep{Szentannai.2019MimosaNet}.
Future work could focus on how to integrate watermarking within such active defense strategies against model stealing.

Furthermore, most current watermarking schemes apply solely to image data, so far.
Only few exceptions, \eg \citep{Jia.2020Entangled}, have proven the applicability of their schemes to other data types.
Future work will have to focus on examining the generality and universal applicability of existing schemes, and, if necessary, their adaptation or extension.

Moreover, most watermarking approaches proposed so far also apply solely to classification tasks.
There exist only few works on watermarking in other ML domains, like reinforcement learning, \eg \citep{Behzadan.2019Sequential}, and data generation with GANs \citep{ong2021protecting}, or image captioning \citep{lim2020protect}.
Therefore, the development of watermarking schemes for other ML applications represents a promising future challenge.

Additionally, so far, existing watermarking schemes are mainly applied and evaluated on rather small research datasets, like MNIST \citep{LeCun.2010MNIST}.
Therefore, the question of their scalability remains open.
Approaches that require training with up to double the initial amount of data might, hence, not be applicable to every scenario.
Thus, future work should assess the practical applicability of existing watermarking schemes to larger real-world datasets and analyze whether the properties they exhibit on the research datasets (efficiency of training, reasonable trigger dataset size, integrity, \etc) hold.

Finally, once that watermarking schemes meet all the technical requirements, another challenge will lie in their adaptation to real-world workflows.
Especially the juridical and organizational workflows will have to be adapted in order to enable asserting ownership claims based on the watermarks.

\section{Concluding Remarks}
\label{sec:conclusion}
Nowadays, ML is used in an increasing number of domains.
With growing complexity of the applied models, employing watermarks to protect intellectual property incorporated in those models has become a major focus both in academia and industry.
This systematic review provided a framework for articulating a comprehensive view on different watermarking schemes.
It introduced a taxonomy to classify and compare existing methods, presented a systematization of the requirements for, and attacks against watermarking schemes, and formulated a unified threat model. 
Guided by the taxonomy, relevant prior research was surveyed.
This work can serve as a solid foundation for analyzing existing watermarking methods, designing new ones, or choosing adequate solutions to given scenarios.
Therefore, it can be used as a reference for researchers and ML practitioners over all domains.

\section*{Acknowledgments}
We acknowledge support by the Open Access Publication Initiative of Freie Universität Berlin.
Additionally, we would like to thank Hengrui Jia for the fruitful discussion about the topic and his feedback on the manuscript.

\bibliographystyle{abbrvnat} 
\bibliography{library}

\end{document}